\documentclass[aps,prl,showpacs,floatfix, twocolumn]{revtex4}
\usepackage{graphicx}
\usepackage{amsfonts}
\usepackage{amssymb}
\begin{document}
\title{Nonlocal Modulation of Entangled Photons}
\author{S. E. Harris}
\email{seharris@stanford.edu}

\affiliation{Edward L. Ginzton Laboratory, Stanford
University, Stanford, California 94305}

\setlength{\abovecaptionskip}{5pt}
\setlength{\belowcaptionskip}{5pt}
\setlength{\intextsep}{5pt}
\setlength{\textfloatsep}{5pt}
\setlength{\floatsep}{5pt}
\setlength{\dbltextfloatsep}{5pt}
\setlength{\dblfloatsep}{5pt}

\date{\today}
\begin{abstract}
We consider ramifications of the use of high speed light modulators to questions of correlation and measurement of time-energy entangled photons.  Using phase modulators, we find that temporal modulation of one photon of an entangled pair, as measured by correlation in the frequency domain, may be negated or enhanced by modulation of  the second photon.  Using amplitude modulators we describe a Fourier technique for measurement of biphoton wave functions with slow detectors. 
 \end{abstract}
\maketitle

\begin{figure}[tbp]
\begin{center}
\includegraphics*[width=3.2truein]{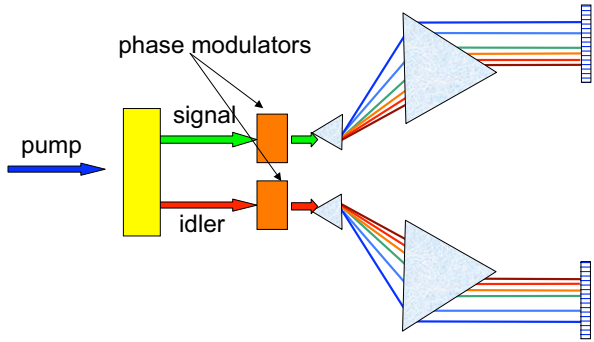}
\end{center}
\caption{(color online). Quantum modulation. Signal and idler photons are phase modulated at the same frequency and with controllable phase. The signal and idler beams are diffracted to linear arrays and the positions of the detected photons on the photodetectors are correlated.}
\label{schematic}
\end{figure} 

It is the intent of this Letter to explore the ramifications of the use of high speed  light modulators to questions of correlation and measurement of time-energy entangled photons \cite{Shih; review}.  We start by considering the system of Fig. \ref{schematic} where a monochromatic pump generates non-degenerate time-energy entangled photon pairs. The signal and idler photons pass through sinusoidal phase modulators. These modulators are driven at the same modulation frequency and are connected by a cable such that their relative phase may be varied. After passing through the modulators the signal and idler photons are dispersed, for example by a prism, and the relative positions of the signal and idler photons are correlated. When the modulation frequency is small as compared to the spectral bandwidth of the signal or idler, we find a consequence of time energy entanglement that we term as nonlocal modulation. Specifically, the modulators act cumulatively  to determine the apparent  modulation depth.  For ideal phase modulators the depth of modulation is the sum of the modulation depths of the signal and idler modulators. When the modulators are run with the same phase, the modulation depths add, when they are run in phase opposition, the modulation depths subtract; two distant modulators with the same modulation depth and opposite phase will have the same (delta function) frequency correlation as when both modulators are absent. 

In the latter portion of this Letter, (Fig. 5), we consider synchronously driven amplitude modulators in both channels. We show how they may be used to measure the waveform of biphotons that are too short to be measured by present day photo-detectors, but are not so short as to be out of the range of high speed (60 GHz) light modulators.

Nonlocal modulation as described here may be thought of as the time-frequency counterpart  of nonlocal dispersion compensation as suggested by Franson \cite{Franson}, and developed experimentally by Shih \cite{Shih-dispersive} and Silberberg \cite{Silberberg-coherentcontrol,Silberberg-temporalshaping} and colleagues. Both phenomena depend on the quantum mechanical addition of probability amplitudes, and both do not have a classical analog.
There is considerable theoretical and experimental work that is pertinent to the work described here. Examples include quantum interference effects \cite{Zou}, nonlocal pulse shaping \cite{Arecchi}, descriptions of time-energy entanglement \cite{Howell}, experiments showing higher-dimensional entanglement using trains of pumping pulses \cite{Gisin}, and experimental  control of the joint spectrum of down-converted photons \cite {Torres,Walmsley}. The waveform measurement technique described below is in part motivated by recent demonstrations of EIT based techniques for generating paired photons that are many cycles long at the modulation frequency \cite{Balic, Kolchin, Vuletic}.

We  develop the theory in the Heisenberg picture where the output of the parametric down-converter is described by frequency domain operators $a_s(\omega)$ and $ a_i^\dagger(\omega_i)$, with 
$\omega_i=\omega_p-\omega$, where $\omega_p $ is the frequency of the monochromatic pump. These operators are expressed in terms of the vacuum fields at the input of the nonlinear generating crystal $a_s(\omega, 0)$ and $a_i^\dagger(\omega_i, 0)$ by frequency dependent functions $A(\omega), B(\omega), C(\omega)$ and $D(\omega)$, all expressed in terms of the signal frequency $\omega$. Thus

\begin{eqnarray}
a_s(\omega)&=A(\omega)\  a_s(\omega, 0)+B(\omega)\  a_i^\dagger(\omega_i, 0)\nonumber\\
a_i^\dagger (\omega_{i})&=C(\omega)\  a_s(\omega, 0)+D(\omega)\  a_i^\dagger(\omega_{i}, 0)\label{solution}. 
\end{eqnarray}
The frequency domain operators are the Fourier transform of time domain operators, i.e.,  $a(\omega)=1/(2\pi) \int_{- \infty}^{ \infty} a(t) \exp(i \omega t) dt $, that in turn have the commutator $[a_j(t_1, 0),a_k^\dagger(t_2, 0)]= \delta(t_1-t_2)\delta_{j,k}$.  The operator $a(t)$ is  normalized so that the paired count rate for a single transverse mode is $R=\langle a^{\dagger}(t) a(t)\rangle$. The coefficients are related by the unitary conditions $A(\omega)C^*(\omega)=B(\omega)D^*(\omega), |A(\omega)|^2-|B(\omega)|^2=1$, and  $|D(\omega)|^2-|C(\omega)|^2=1$.
Gatti et al. \cite{Gatti,Gatti2} have shown  that for low parametric gain where there is, most often,  only a single photon at both the signal and idler frequencies, and with $|A(\omega)|=1$, the biphoton is described by the function $\phi(\omega)=A(\omega)C^*(\omega)$. 

\begin{figure}[tbp]
\begin{center}
\includegraphics*[width=3.2truein]{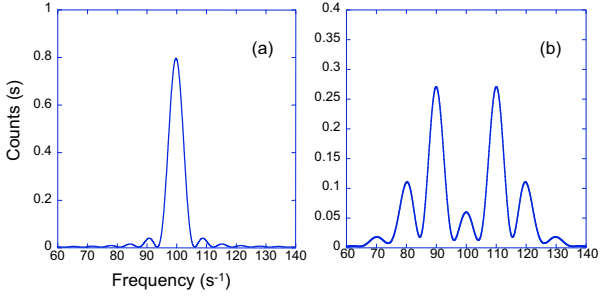}
\end{center}
\caption{(color online). Spectrum of the signal frequency after passing through the modulator. In part (a) the modulation frequency $\omega_m=0.1$ and is small as compared to the linewidth of one unit of the biphoton. In part (b) $\omega_m=10$ and is large as compared this linewidth. Counts are per bandwidth in a gatewidth $\rm{T}$.} \label{Spectra}
\end{figure}

The phase modulators in the signal and idler channels are assumed to be periodic with the same modulation frequency $\omega_m $ and with Fourier series  $\sum_{n} q_n \exp(-i n \omega_m t)$, and $\sum_{m} r_m \exp(-i m \omega_m t)$, respectively. In the frequency domain these modulators are described by
\begin{eqnarray}
m_s(\omega)=\sum_{n=-\infty}^{\infty} q_n \delta \left(\omega-n \omega_m\right)\nonumber\\
m_i(\omega)= \sum_{m=-\infty}^{\infty} r_m \delta \left(\omega-m \omega_m\right).
\label{modulators} 
\end{eqnarray}

Time varying modulators multiply the incoming temporal waveform, and their transforms convolve with the Fourier transform of this waveform.  The operators at the output of the signal and idler modulators are obtained by combining Eqs. (\ref{solution}) and Eqs. (\ref{modulators}) and are 
\begin{eqnarray}
a_s(\omega)=\sum_{n=-\infty}^{\infty} q_n A(\omega-n \omega_m) a_s[(\omega-n \omega_m),0]+\nonumber \\ q_nB(\omega-n \omega_m) a_i^\dagger[ (\omega_i+n \omega_m),0] \nonumber \\
a_i^\dagger(\omega_{i})=\sum_{m=-\infty}^{\infty} r_m C(\omega-m \omega_m) a_s[(\omega-m \omega_m),0]+\nonumber \\ r_m D(\omega-m \omega_m) a_i^\dagger [(\omega_i+m \omega_m),0] \nonumber\\
\label{output} 
\end{eqnarray}

With the modulators present, the spectrum at the signal and idler frequencies as would be observed as a function of position on the upper and lower screens of Fig (1) are $S(\omega)= (2 \pi) \langle a_s^\dagger(\omega) a_s(\omega)\rangle$, and $I(\omega)= (2 \pi) \langle a_i^\dagger(\omega) a_i(\omega)\rangle$, and evaluate to
\begin{eqnarray}
S(\omega)=\frac{\rm{T}}{2 \pi} \sum_{n=-\infty}^{\infty} |q_n|^2 |B(\omega-n \omega_m)|^2\nonumber\\
I(\omega)=\frac{\rm{T}}{2 \pi} \sum_{m=-\infty}^{\infty} |r_m|^2 |C(\omega+m \omega_m)|^2
\label{spectrum} 
\end{eqnarray}
The total number of counts at either frequency in gatewidth $\rm{T}$ is obtained by integrating Eq.(\ref{spectrum}) over $\omega$.

We take the coefficients of Eq.(\ref{solution}) to correspond to a rectangular wavefunction with a temporal width equal to one unit \cite{Rubin} . These are $A(\omega)=D(\omega)=1$, and $B(\omega)=C^{*}(\omega)=\exp(-i x) \sin(x)/x$, where $x=(\omega-\omega_0)/2$ and $\omega_0$ is the center frequency of the signal spectrum.

We assume ideal phase modulators with the functional form  $\exp[i \delta \sin\omega_m t]$ at the signal and idler frequencies. The modulator coefficients in Eq.(\ref{modulators}) are then Bessel functions with $q_n=J_{n}(-\delta_s)$, and $r_m=J_{m}(-\delta_i)$. Fig. 2 shows the spectrum of the signal frequency after passing through the modulator. Two limiting cases are of interest; in the first, Fig. 2 (a), the modulation frequency $\omega_m=0.1$ is small as compared to the linewidth of the biphoton ($\rm{1}$ unit), and the spectrum is nearly the same as without the modulator present. In the second case, Fig. 2 (b), the modulation frequency $\omega_m=10$ is large as compared to the linewidth of the biphoton. In both cases we take the modulation depth $\delta_{s}=2$ radians. In all cases with ideal phase modulators the rate of generation of paired photons is $R=1/(2\pi)\int_{-\infty}^{\infty} |B(\omega)|^2 d\omega$.

To measure the two frequency correlation function one would observe a signal ``click" on one screen, and rather immediately an idler click on the other screen. The position of the idler  click relative to the position $\omega_p-\omega_s$ is recorded and integrated over all positions at the signal frequency.  

We take the two frequency correlation function, as observed in the frequency domain as $G^{(2)}(\omega_s,\omega_i )=(2 \pi)^{2} \langle a_i^\dagger(\omega_i) a_s^{\dagger}(\omega_s) a_s(\omega_s)  a_i(\omega_i)  \rangle$. We expand by Wicks theorem and find that the result is the sum of two terms. The first term is the same as that obtained by a classical (frequency domain) correlation of the intensities of the signal and idler. The second term is the result of time energy entanglement, and as described below exhibits what we term as nonlocal modulation. With a relative detuning between the signal and the idler defined as $\Delta=\left[\omega_p-(\omega_s+\omega_i)\right]$, and with $\rm{T}$ as the gate width, the classical term $c(\Delta)$ is
\begin{eqnarray}
c(\Delta)=\left(\frac{\rm{T}}{2\pi}\right)^2\int_{-\infty}^{\infty} \sum_{n=-\infty}^{\infty}\sum_{m=-\infty}^{\infty}|q_n|^2 |r_m|^2 \times\nonumber \\ 
|B(\omega-n \omega_m)|^2 |C[\omega+\Delta+m \omega_m]|^2 d\omega
\label{classical}
\end{eqnarray}
This classical correlation function is equal to the convolution of the signal and idler spectra of 
Eq.(\ref{spectrum}), is non-zero for all $\Delta$, and is the same as that  which would be obtained using any light source with the same spectrum. 

While the classical correlation function is continuous, the quantum portion is not; instead it is a set of delta functions that are located at integer multiples of the modulation frequency $\omega_m$.
With $\rm{z}=(\Delta/\omega_m)$, the quantum term $q(\Delta)$ evaluates to 
\begin{eqnarray}
&q(\Delta)&=\left(\frac{\rm{T}}{2\pi}\right) \sum_{z=-\infty}^{\infty} f \left(z\right) \delta[\Delta+z\omega_m]  \\
&f(z)=&\int_{-\infty}^{\infty}\left|\sum_{n=-\infty}^{\infty}q_{n}^{*} r^{*}_{z-n} C(\omega-n \omega_m)
A^{*}(\omega-n \omega_m)\right|^2 d\omega \nonumber
\label{quantum}
\end{eqnarray}

If both modulators are absent, then Eq.(6) yields only a single delta function at zero. If a modulator is present in only one channel, the quantum term is a comb of delta functions.  With modulators in both channels, and when the modulation frequency is small as compared to the spectral width of the biphoton wavepacket, the modulators interact in the sense that the cumulative modulation depth of the biphoton depends on the sum of the modulation depths of the independent modulators.  
\begin{figure}[tbp]
\begin{center}
\includegraphics*[width=3.4truein]{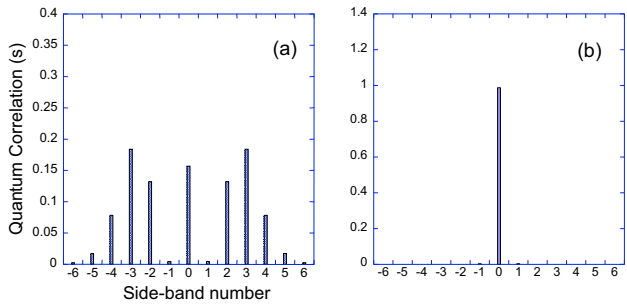}
\end{center}
\caption{Quantum correlation versus sideband number. (a) The signal and idler are modulated with the same phase so that $\delta_s= \delta_i=2$.  Here the modulators act cumulatively. (b) The signal and idler are modulated with opposing phases so that $\delta_s=2$ and $ \delta_i=-2$. Here, the modulators negate each other. In both portions of this Figure, $\omega_m=0.1$, as compared to the non-modulated biphoton bandwidth of unity.}
\label{nonlocal}
\end{figure} 

Figure \ref{nonlocal} shows the nonlocal correlation obtained with both modulators on. In Fig. \ref{nonlocal} (a), the signal and idler channels are modulated with the same phase so that $\delta_s=\delta_i=2 $.  The modulators act cumulatively, so as to produce the same correlation as would a single modulator with a modulation depth of four. In Fig. \ref{nonlocal} (b), the modulators are driven with opposing phases, i.e. $\delta_s=2$ and  $\delta_i=-2$  . Here the modulators negate each other so that the effective biphoton modulation depth is zero. 

Of importance, the frequency of modulation in Fig.~ \ref{nonlocal} is small $(\omega_m=0.1)$ compared to the bandwidth of the biphoton which is unity. Fig. 4 shows the correlation with the same conditions as Fig. \ref{nonlocal}, but this time with $(\omega_m=10)$. Here, we find that the nonlocal modulation effects disappear. The requirement for cumulative modulation is that the modulation frequency be sufficiently low that all portions of the initial biphoton wavepacket experience nearly the same phase. Since nonlocal modulation is the time-frequency counterpart of nonlocal dispersion compensation \cite{Franson}, the comparable requirement for  nonlocal dispersion compensation is that the inverse of the rate of change of the dispersive constant with frequency be small as compared to the width of the frequency domain correlation function.  

\begin{figure}[tbp]
\begin{center}
\includegraphics*[width=3.4truein]{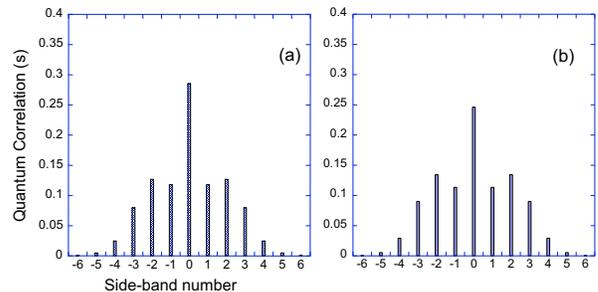}
\end{center}
\caption{Here, the modulation frequency $\omega_m=10$ is large as compared to the biphoton bandwidth of unity. (a) $\delta_s=2$ and  $\delta_i=2$, and (b) $\delta_s=2$ and  $\delta_i=-2$. Cumulative modulation effects are no longer observed.}
\label{large modulation freq}
\end{figure}

For ideal phase modulators, with $\rm{R}$ as the rate of generation of paired photons [Eq.($\ref{spectrum}$)] we obtain the integrals
\begin{eqnarray}
&&\int_{-\infty}^{\infty} c(\Delta)d\Delta=R^{2} T^{2} \nonumber\\
&&\int_{-\infty}^{\infty} q(\Delta)d\Delta=R T
\label{sums}
\end{eqnarray}
The first of these relations follows immediately by integration of Eq.(\ref{classical}). The second requires the assumption that $|A (\omega)|$ is approximately unity, i.e., small parametric gain in the generating crystal.. From Eq.(\ref{sums}), it is clear that to observe the quantum effects described here, that the gate width $\rm{T}$ be sufficiently small that $\rm{R T<<1}$.  

\begin{figure}[h]
\begin{center}
\includegraphics*[width=3.2truein]{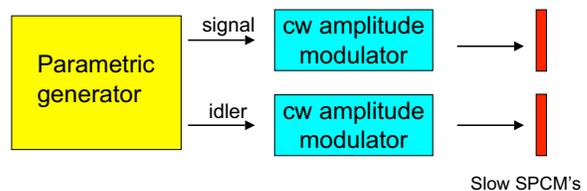}
\end{center}
\caption{Measurement of biphoton wave packets. Amplitude modulators in the signal and idler paths are driven synchronously. The coincidence count rate is measured by slow detectors and is plotted as a function of modulation frequency. The Fourier transform of this plot is the square of the magnitude of the biphoton wavefunction.}
\end{figure} 

We next consider synchronously driven amplitude modulators and describe a Fourier technique for measuring biphoton wavepackets that are shorter than the temporal resolution of existing single photon counting modules. First note that because the incident biphotons arrive at the modulators at random times, the coincidence count rate will be unchanged unless the modulators are driven at the same frequency. We recognize that we may accomplish a time to frequency Fourier transform by multiplying in the time domain, i.e. by modulating, and integrating over time. In Fig. 5 signal and idler photons are amplitude modulated and are incident on detectors whose response time is long as compared to the temporal length of the photon wavepacket, and short as compared to the inverse pair generation rate. The modulation frequency at the signal and the idler is varied and the coincidence count rate is measured as a function of this frequency. With $\delta_s \delta_i<<1$, the inverse Fourier transform of the measured curve is the the square of the magnitude, or equivalently the quantum portion of the Glauber correlation function of the biphoton wavefunction. 

\begin{figure}[h]
\begin{center}
\includegraphics*[width=3.4truein]{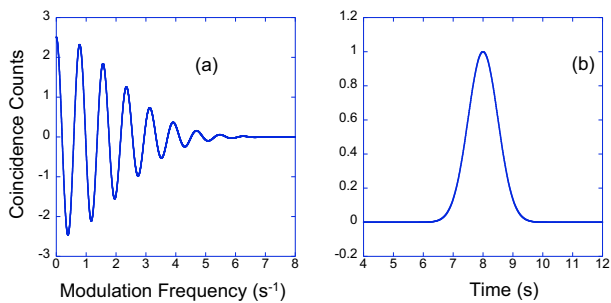}
\end{center}
\caption{Biphoton Wavefunction Measurement (a) Coincidence Counts per bandwidth versus modulation frequency with the average count rate set to zero.  (b) Fourier Transform of part (a).}
\label{FourierFig}
\end{figure} 

The derivation is reasonably straightforward. We begin by multiplying the signal and idler operators at the output of the generating crystal by $1+\delta_s \cos(\omega_m t)$ and  $1+\delta_i \cos(\omega_m t)$, respectively. We form the correlation function $G^{(2)}=\langle a_{i}^{\dagger}(t_1) a_{s}^{\dagger}(t_2) a_{s}(t_2) a_{i}(t_1) \rangle$, Fourier transform and expand by Wicks theorem. Two non-zero terms are obtained: The first is the quantum term and the second is a classical term whose magnitude varies as the square of the generation rate. We take $\tau=t_2-t_1$, average $t_1$ over the modulation period, and integrate over $\tau$.  With the biphoton wavefunction denoted by $\phi(\omega)=A(\omega)C^*(\omega)$, and $\kappa=\delta_s\delta_i/2\pi$, the quantum term to lowest order in $\kappa$ is
\begin{equation}
\mathcal{F}(\omega_m)=\kappa \int_{-\infty}^{\infty}[\phi(\omega+\omega_m) \phi^*(\omega)+cc] d\omega
\label{Fourier}
\end{equation}

Eq. (\ref{Fourier}) is the inverse Fourier transform of the square of the magnitude of the biphoton wavefunction.  Figure \ref{FourierFig}(a) shows the coincidence count rate as a function of the modulation frequency for a Gaussian biphoton wavefunction with a duration of one unit and a delay of eight units in the signal channel.  Fig. \ref{FourierFig}(b) shows the wavefunction obtained by taking the inverse Fourier cosine transform of Fig. \ref{FourierFig}(a). We note that a different method for measuring short biphotons by correlation has been demonstrated by Silberberg and colleagues \cite{Silberberg-temporalshaping}.

The author thanks S. Sensarn, I. A. Khan, S. Du, P. Kolchin and M. Scully for helpful discussions. This work was supported by the Defense Advanced Research Projects Agency, the U.S. Air Force Office of Scientific Research, and the the U.S. Army Research Office.

\end{document}